\documentclass[letter]{achemso}

\usepackage[T1]{fontenc}
\usepackage{amsmath,mathtools}
\usepackage{graphicx}
\usepackage{siunitx}
\usepackage{subcaption}
\usepackage{xcolor}

\DeclareSIUnit\angstrom{\text{Å}}

\newcommand{\codename}[1]{{\sc {#1}}}

\newcommand{\phonopy}{\codename{phonopy}}
\newcommand{\vasp}{\codename{vasp}}

\title{High-Pressure Inelastic Neutron Spectroscopy: A true test of Machine-Learned Interatomic Potential energy landscapes}

\author{Jeff Armstrong}
\email{jeff.armstrong@stfc.ac.uk}
\affiliation{ISIS Neutron and Muon Source, UKRI STFC, Rutherford Appleton Laboratory, Didcot, UK}
\affiliation{Department of Chemistry, University of Bath, Bath, UK}

\author{Adam Jackson}
\affiliation{Scientific Computing, UKRI STFC, Rutherford Appleton Laboratory, Didcot, UK}

\author{Alin Elena}
\affiliation{Scientific Computing, UKRI STFC, Daresbury Laboratory, Daresbury, UK}

\begin{document}

\begin{abstract}
Machine-learned interatomic potentials (MLIPs) promise to provide near density-functional theory accuracy at a fraction of the computational cost, offering a transformative route toward genuinely predictive chemistry. Yet their predictive validity beyond the training regime remains largely untested experimentally. 

Here we use pressure-dependent broadband inelastic neutron spectroscopy (INS) as a direct experimental probe of MLIP transferability. Employing a newly developed high-pressure superalloy clamp cell, we measure INS spectra of crystalline 2,5-diiodothiophene at 10~K under ambient conditions and at 1.5~GPa. A MACE-based MLIP, fine-tuned on targeted DFT data, reproduces the experimental spectra across 0--1200~cm$^{-1}$ at both pressures and remains thermodynamically stable under rigorous molecular dynamics validation at 300~K. The model captures systematic pressure-induced blue shifts arising from steric stiffening and reproduces an anomalous red shift at 453~cm$^{-1}$ driven by pressure-modified intermolecular interactions, providing direct validation of its many-body character. 

This constitutes the first experimental demonstration of MLIP transferability across distinct thermodynamic states using neutron spectroscopy, and establishes high-pressure INS as a stringent benchmark for predictive machine-learned potentials.
\end{abstract}

\maketitle

Machine-learned interatomic potentials (MLIPs) have transformed atomistic modelling by delivering accuracy close to density-functional theory (DFT) at a fraction of the computational cost.\cite{batatiaMACEHigherOrder2022,kovacsEvaluationMACEForce2023} This enables simulations of vibrational response, collective structural fluctuations, diffusive motion, and slow dynamical processes on spatial and temporal scales directly comparable with experiment. As MLIPs become embedded in predictive materials modelling, however, rigorous experimental validation becomes essential. Crucially, agreement with the first-principles data used for training or with static structural observables\cite{Ferretti2025JCIM,Sivaraman2024JPCM,Gurlek2025npjCompMater} does not guarantee that the underlying forces remain accurate when the thermodynamic state is perturbed.

Many technologically relevant materials operate in regimes where weak, cooperative, and environment-dependent interactions dominate their behaviour. In zeolites used for catalysis,\cite{Morton2025Microporous,Morton2025RSC_Sustainability} adsorption and reactivity depend sensitively on collective framework flexibility and subtle many-body interactions between guest molecules and acid sites. In metal--organic frameworks,\cite{Mor2025Langmuir,Bahadur2025JPCL} gas uptake and selectivity emerge from correlated host--guest motion, transient pore distortions, and crowding under confinement. In electrolytes,\cite{Shigenobu2023JPCB} ion transport arises from coupled rotational and translational dynamics within fluctuating coordination environments. Organic electronic materials\cite{MarinVilla2025JPCL,Druzbicki2021JPCL,Schweicher2019,Banks2023} rely on delicate balances between intermolecular packing, low-energy phonons, and dynamic disorder to govern charge mobility. In all such systems, many-body forces govern the accurate description of rotational barriers, diffusion coefficients, vibrational properties, and ultimately the macroscopic transport behaviour of the material.

Broadband inelastic neutron spectroscopy (INS) provides a direct experimental probe of these forces because vibrational frequencies correspond to second derivatives of the potential-energy surface, and simulated neutron-weighted spectra can be compared with experiment in a one-to-one manner.\cite{Dymkowski2018AbINS,Armstrong2022JPhysCommun} Accurate reproduction of INS spectra therefore tests not only equilibrium structure but also the curvature and coupling of the underlying many-body force landscape.

Crucially, a fundamental limitation of INS is that measurements must typically be performed at cryogenic temperatures, constraining temperature as a thermodynamic axis for validating model transferability. This presents a serious challenge, because without demonstrated transferability, the central benefit of MLIPs is eroded, given that their training cost can approach that of ab initio methods. To overcome this central limitation, we recast pressure as a deliberate thermodynamic axis of validation, enabling stringent tests of transferability without sacrificing the sharp spectral resolution of 10 K INS measurements. However, high-pressure INS presents formidable experimental challenges, stemming from the limited sample volumes and significant parasitic scattering associated with bulky pressure cells. Recent advances on the TOSCA spectrometer have mitigated these constraints through the deployment of an ultra-tough NiCrAl alloy cell with intrinsically low neutronic background, enabling robust measurements at gigapascal pressures\cite{Pinna2018NeutronGuideTOSCA,armstrongpressure}. This capability allows vibrational spectra to be recorded under systematically compressed intermolecular separations within the same crystalline material, probing modified interatomic force constants and providing a stringent experimental test of MLIP transferability across thermodynamic conditions.

\begin{figure}[ht!]
 \centering
 \includegraphics[width=0.85\linewidth]{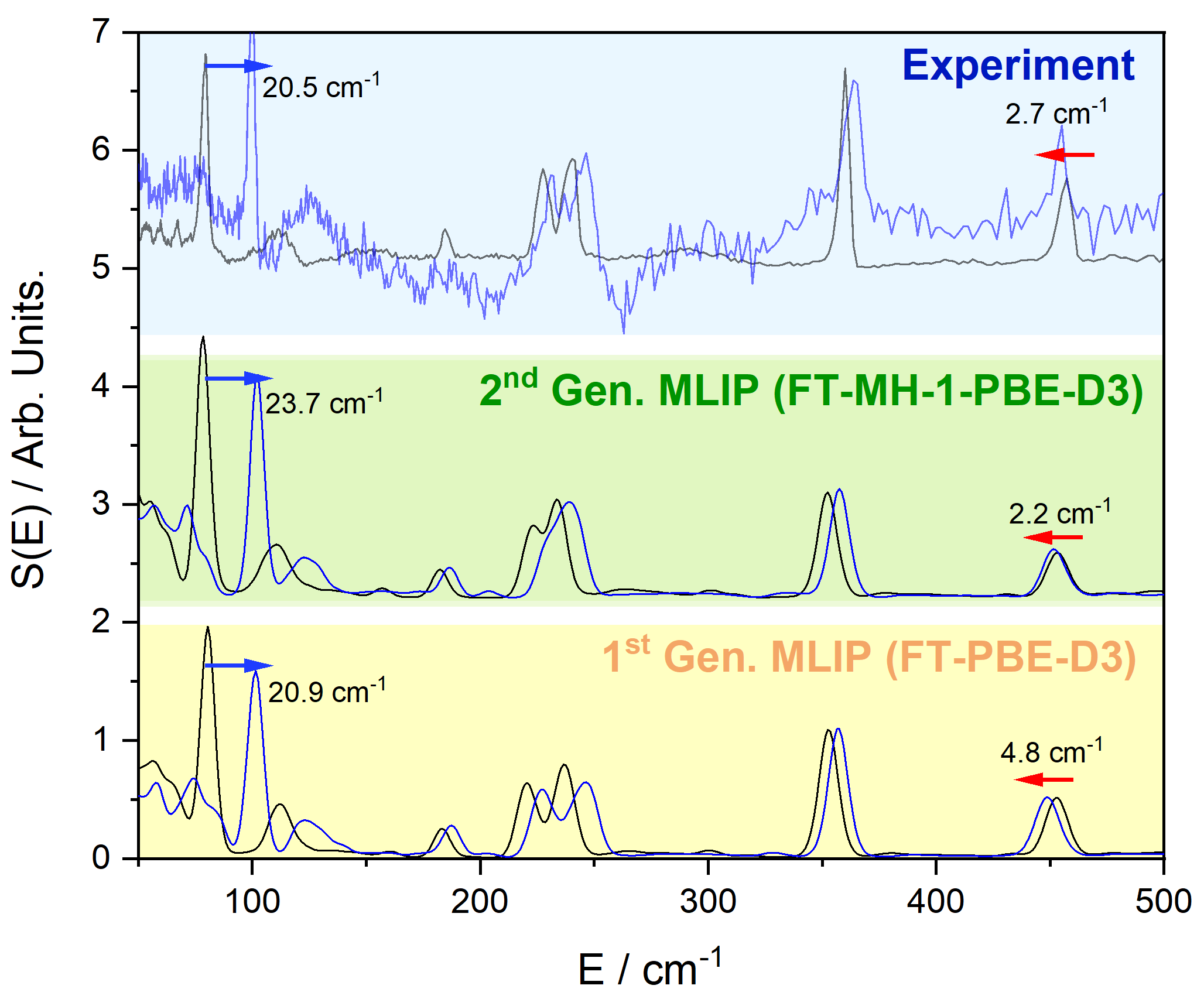}
 \caption{Experimental INS spectra (top blue panel) compared with MLIP-derived neutron-weighted spectra for 1$^{\text{st}}$ and 2$^{\text{nd}}$ generation models (yellow and green panels respectively) at ambient pressure (black lines) and 1.5~GPa (blue lines). Illustrative blue and red shifts are shown as colored arrows.}
 \label{exptsimcompare}
\end{figure}

As a model system we examine crystalline 2,5-diiodothiophene, whose well-defined INS
line shapes and fine dispersion character make it a sensitive
probe of subtle changes in the potential-energy surface under compression. The measured INS spectra of crystalline 2,5-diiodothiophene at ambient pressure and 1.5~GPa retain clearly resolved vibrational features across 0–1200~cm$^{-1}$ following empty-cell subtraction. Representative subtraction spectra are provided in the Supporting Information. The dominant pressure response occurs below 500~cm$^{-1}$ (Figure~\ref{exptsimcompare}), where multiple bands shift to higher energy under compression, consistent with steric stiffening as intermolecular separations decrease. In contrast, a feature near 453~cm$^{-1}$ exhibits a reproducible red shift. This mode has previously been assigned to an out-of-plane ring deformation coordinate,\cite{Parker2017JPCC} demonstrating that compression does not uniformly harden all vibrational degrees of freedom, but can selectively soften specific internal motions through subtle changes in intermolecular coupling.

To construct a transferable MLIP suitable for pressure-dependent validation, a first-principles reference description of the potential-energy surface is required. Density functional theory (DFT) calculations were therefore performed using the projector-augmented wave method implemented in \vasp{} to generate an initial PBE-D3 training dataset
\cite{kresseEfficientIterativeSchemes1996,kresseUltrasoftPseudopotentialsProjector1999,perdewGeneralizedGradientApproximation1996,grimme2010,grimmeEffectDampingFunction2011}. These calculations provided energies, forces, and stresses for pressure-optimised supercells that span both ambient and compressed configurations.
The system-specific MLIP was then developed by fine-tuning the MACE-MP-0 foundation model on this data\cite{batatiaFoundationModelAtomistic2024}. Rather than relying solely on small harmonic displacements about equilibrium, the training set was iteratively enriched with configurations sampling local strain, anisotropic lattice distortions, equation of state like structures, geometry optimisation and finite-temperature fluctuations via molecular dynamics using ASE\cite{Larsen2017} and janus\_core\cite{kasoar_2026_18713710}. 
For the exact details of the of the finetuning configuration selection see the Supplementary Information. This refinement strategy exposes the model to regions of configuration space relevant to both thermodynamic states. Such training is particularly demanding for molecular crystals, where weak intermolecular dispersion, steric contacts, and subtle pressure-induced reorientation of the herringbone packing must be captured consistently in order to preserve both structural stability and the correct curvature of the potential-energy surface.
Additional models were generated by fine-tuning ``second generation'' multi-headed MACE foundation models\cite{batatia2025crosslearningelectronicstructure} with the same training configurations and with different density-functional approximations, PBE-D3\cite{perdewGeneralizedGradientApproximation1996,grimme2010,grimmeEffectDampingFunction2011} and R2SCAN-D4.\cite{furnessAccurateNumericallyEfficient2020,caldeweyherGenerallyApplicableAtomiccharge2019,ehlertR2SCAND4DispersionCorrected2021}. Full details of these models are contained in the Supporting Information.

TOSCA spectra were simulated from the MLIP using \phonopy{} and \codename{abins}\cite{togoFirstprinciplesPhononCalculations2023,Dymkowski2018AbINS}. 
Figure~\ref{exptsimcompare} compares the experimental spectra with the simulated spectra from the first and second generation fine-tuned MLIPs (PBE based). Crucially, these calculations can be run in minutes on a personal computer, whereas equivalent DFT calculations typically require days on high-performance computing resources. Agreement is near-quantitative across the full measured energy range, with the first and second generation models each performing slightly better in different regions of the spectra. Strikingly, both generations of MLIP reproduce the pressure response of the vibrational modes, including the anomalous red shift at 453~cm$^{-1}$. Because the low-energy region is governed by weak intermolecular interactions and subtle packing rearrangements under compression, accurate reproduction of these pressure-induced shifts demonstrates that the MLIPs correctly describe how the curvature of the intermolecular potential-energy surface evolves with pressure.

\begin{figure}[ht!]
 \centering
 \includegraphics[width=0.85\linewidth]{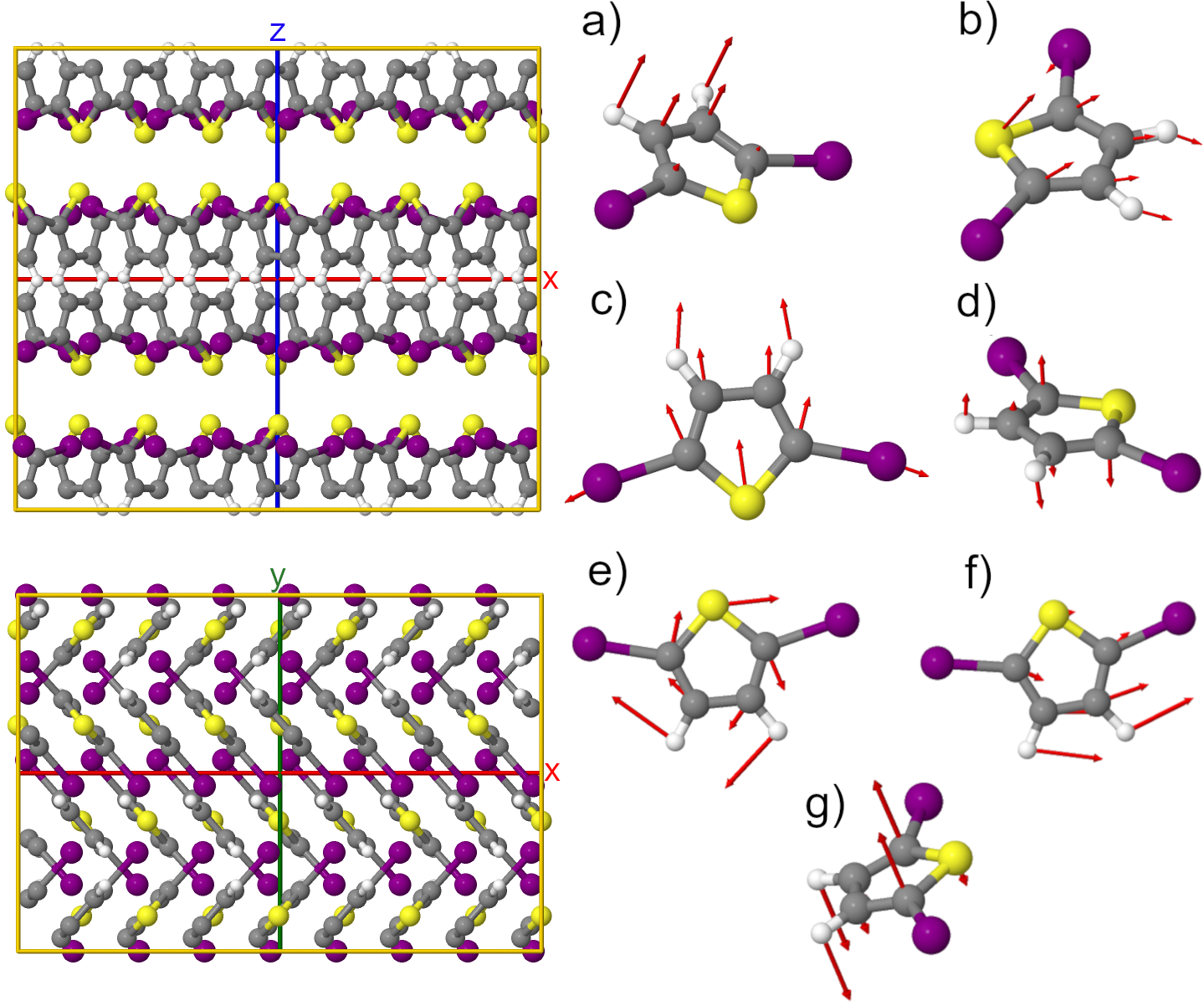}
 \caption{Left: MLIP-relaxed packing motif for the MLIP, illustrating bilayer formation and herringbone arrangement. Right: Molecular depictions/assignments of vibrational modes across the 80-500~cm$^{-1}$ range in ascending frequency.}
 \label{struct}
\end{figure}

Analysis of the MLIP-relaxed structures reveals that the crystal comprises halogen-bond-mediated bilayers stacked along the $z$-axis and a distorted herringbone packing motif within the $x$--$y$ plane (Figure~\ref{struct}, left panel). Compression to 1.5~GPa produces pronounced anisotropic lattice contraction, with a 4.40\% reduction along the $z$-axis (parallel to the halogen-bonded bilayers) and a 2.73\% contraction along the $x$-axis. This anisotropy is accommodated by a pressure-induced adjustment of the herringbone tilt angles (Figure~\ref{struct}, left panel), preserving halogen contacts while modifying intermolecular separations and relative molecular orientation.
The relative orientation of neighboring rings is altered by \SI{3}{\degree} under compression from \SIrange{0}{1.5}{\giga\pascal}.

The vibrational mode visualisations in Figure~\ref{struct} (right panel) clarify how these structural changes translate into spectral shifts. The modes that blue shift under compression (Figure~\ref{struct}~a-f) are dominated by collective displacements of the rigid ring motif relative to neighbouring molecules. In these cases, the internal geometry of the five-membered ring remains essentially intact; the motion primarily alters intermolecular contact distances. Under compression, such displacements increase steric repulsion between adjacent motifs, steepening the local energy curvature and naturally producing systematic blue shifts. By contrast, the 453~cm$^{-1}$ mode (Figure~\ref{struct}~g) corresponds to an internal out-of-plane deformation of the ring framework itself. Through the tilting adjustment to the herringbone packing the associated ring--ring electronic overlap changes, reducing the effective restoring force along this coordinate and providing a structural explanation for the observed red shift.

Although the ability to rapidly compute phonon spectra, and therefore any macroscopic properties derived from them, across a wide range of pressures is already highly useful, the ultimate value of an MLIP lies in its ability to describe thermally activated motion and collective behaviour of functional materials at finite temperature. Ion migration in solid electrolytes, molecular diffusion in porous frameworks, lattice thermal transport, mechanical response, and thermodynamic observables all emerge from finite-temperature sampling of the underlying energy landscape. A model that reproduces low-temperature vibrational spectra but exhibits instability or unphysical behaviour under thermal sampling would therefore have limited predictive utility. It is essential to demonstrate that the same potential behaves sensibly in the regime from which macroscopic transport and structural properties are derived.

While both generations of MLIP reproduce the vibrational spectra with comparable accuracy, the two models differ in how they incorporate system-specific training data. The first-generation model was obtained through conventional fine-tuning of the MACE--MP--0 foundation model, in which the network weights are directly updated to reproduce the targeted training configurations. Although this approach is technically straightforward and lowers the barrier to constructing MLIPs, it can in principle lead to \textit{catastrophic forgetting}, whereby aspects of the broader chemical knowledge encoded in the pretrained model are overwritten during fine-tuning. This limitation may become important when the model is used in molecular dynamics simulations, where the system can explore configurations that lie outside the relatively narrow region of configuration space represented in the fine-tuning dataset.

By contrast, second-generation foundation models are trained on a much larger dataset of approximately $10^{8}$ atomic configurations and employ a multi-head architecture designed to mitigate catastrophic forgetting. This framework allows system-specific refinement while preserving the broader representation of chemical environments learned during pretraining, providing a more robust description when exploring configurations away from the harmonic neighbourhood of the training structures. For this reason, the finite-temperature molecular dynamics simulations described below were performed using the second-generation MLIP.

\begin{figure}[ht!]
 \centering
 \includegraphics[width=0.85\linewidth]{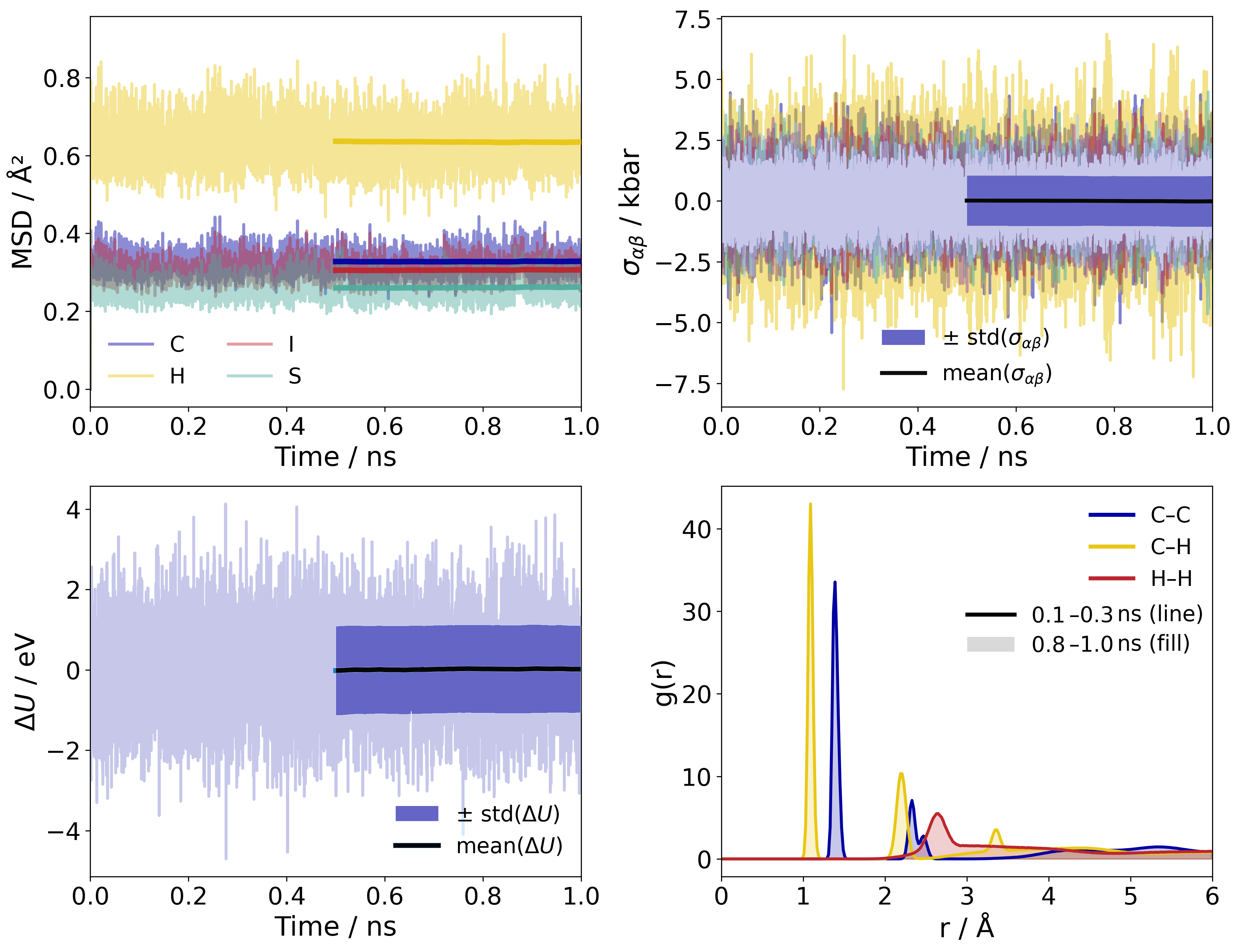}
\caption{
Validation of the 2\textsuperscript{nd} Gen. MLIP (FT-MH-1-PBE-D3) simulation at 300~K. 
Top left: The mean-squared displacement of each atomic component. 
Top right: Fluctuations of the components of the stress tensor. 
Bottom left: Fluctuation of the potential energy. 
1000-point running averages are shown for each plot as dark lines, and std are shown as shaded windows as a guide for the eye. 
Bottom right: Radial distribution functions for C--C, H--H, and C--H, sampled between 0.1--0.3~ps (solid lines) and between 0.9--1.0~ps (shaded plots).
}
 \label{moreMLIPcomp}
\end{figure}

The MD was performed at 300~K in a 1152-atom supercell (Figure~\ref{moreMLIPcomp}) using LAMMPS and the MLIAP package\cite{LAMMPS}. Over nanosecond timescales, the crystal structure remains intact with no evidence of drift or incipient instability. The mean-squared displacements of all atomic species plateau at physically reasonable values, indicating bounded vibrational and librational motion rather than structural disorder. The components of the stress tensor exhibit stationary fluctuations about zero applied pressure, reflecting mechanical stability of the condensed phase. In equilibrium statistical mechanics, stress autocorrelation functions are directly related to shear viscosity and mechanical response via Green--Kubo formalisms, so the absence of systematic drift or anomalous variance indicates a mechanically coherent description under thermal sampling. Radial distribution functions computed over early and late trajectory windows are indistinguishable, confirming preservation of local bonding environments and intermolecular packing. Because the radial distribution function underpins structural observables accessible through diffraction, its invariance demonstrates structural consistency of the simulated ensemble. The potential energy likewise displays bounded, stationary fluctuations about a well-defined mean. In the canonical ensemble, energy fluctuations are directly related to the heat capacity through the fluctuation--dissipation relation, $C_V \propto (\langle E^2 \rangle - \langle E \rangle^2)/T^2$, so the absence of drift or anomalous growth indicates a finite and physically meaningful thermodynamic response.

These results do not constitute direct experimental validation at 300~K; rather, they demonstrate that the microscopic quantities from which macroscopic transport, mechanical, and thermodynamic properties are derived behave sensibly under finite-temperature sampling. The agreement with pressure-dependent INS is therefore embedded within a thermodynamically stable and mechanically consistent description of the system, supporting the use of the MLIP for predictive simulation of dynamical and transport phenomena in molecular materials.

As a final point, we highlight that the refinement protocol appears to be surprisingly robust, with the additional fine tuned MLIP from R2SCAN-D4 DFT data also producing accurate comparison to experimental phonons (See Supplementary Information). This combined with the fact we used three different short range foundation models (mace-mp-0b3, mace-omat-0\cite{batatiaFoundationModelAtomistic2024} and mh-1\cite{batatia2025crosslearningelectronicstructure}), furthers this case. We also tested the very recently developed electrostatic architectures via MACE-POLAR-1-L\cite{batatia2026macepolar1polarisableelectrostaticfoundation}, which despite inclusion of electrostatic long range forces, appears to still need fine tuning to describe experiment. Details of these additional calculations are provided in the Supplementary Information; however, the central conclusion is that the ultimate predictive accuracy remains governed by the quality of the underlying exchange--correlation functional.

In summary, pressure-dependent INS establishes a direct and experimentally grounded route for validating machine-learned interatomic potentials across distinct thermodynamic states. By perturbing intermolecular separations while retaining cryogenic spectral resolution, we probe not only equilibrium vibrational structure but the pressure-dependent reshaping of the underlying potential-energy surface. A single MACE-based MLIP reproduces the measured spectra at ambient pressure and 1.5~GPa, capturing both steric-driven blue shifts and a subtle pressure-induced red shift arising from modified intermolecular electronic interactions. 

Crucially, this agreement in the harmonic regime is embedded within a thermodynamically stable and mechanically consistent description at 300~K, where the microscopic quantities that govern diffusion, viscosity, structural correlations, and heat capacity behave sensibly under finite-temperature sampling. The model therefore supports extraction of macroscopic transport coefficients, collective dynamical observables, and structural response functions from a single transferable representation of the energy landscape. 

Pressure-dependent INS thus provides more than a spectroscopic benchmark; it furnishes a framework for experimentally anchoring MLIPs in a manner directly relevant to functional materials. When validated in this way, MLIPs become not merely interpolative force fields, but predictive tools for modelling transport, thermodynamic response, and structure–dynamics relationships in molecular and condensed systems.

\section*{Data Availability}
Upon publication all data will be made available on a public repository.

\section*{Acknowledgements}
A.M.E. was supported by the Ada Lovelace Centre at the Science and Technology Facilities Council (https://adalovelacecentre.ac.uk/), the Physical Sciences Data Infrastructure (https://psdi.ac.uk; jointly STFC and the University of Southampton) under grants EP/X032663/1 and EP/X032701/1, and EPSRC under grants EP/W026775/1 and EP/V028537/1.
We are grateful for computational
support from the Scientific Computing Department Science and Technology Facilities Council via scarf high performance computing facility(scarf.rl.ac.uk).

\bibliography{bibliography}

@article{Dymkowski2018AbINS,
  title        = {AbINS: The modern software for {INS} interpretation},
  author       = {Dymkowski, Krzysztof and Parker, Stewart F. and Fern{\'a}ndez-Alonso, Felix and Mukhopadhyay, Sanghamitra},
  journal      = {Physica B: Condensed Matter},
  volume       = {551},
  pages        = {443--448},
  year         = {2018},
  doi          = {10.1016/j.physb.2018.02.034},
  url          = {https://www.sciencedirect.com/science/article/pii/S0921452618301546}
}

@article{Pinna2018NeutronGuideTOSCA,
  title        = {The neutron guide upgrade of the TOSCA spectrometer},
  author       = {Pinna, Roberto S. and Rudi{\'c}, Svemir and Parker, Stewart F. and Armstrong, Jeff and Zanetti, Matteo and {\v S}koro, Goran and Waller, Simon P. and Zacek, Daniel and Smith, Clive A. and Capstick, Matthew J. and McPhail, David J. and Pooley, Daniel E. and Howells, Gareth D. and Gorini, Giuseppe and Fernandez-Alonso, Felix},
  journal      = {Nuclear Instruments and Methods in Physics Research Section A: Accelerators, Spectrometers, Detectors and Associated Equipment},
  volume       = {896},
  pages        = {68--74},
  year         = {2018},
  doi          = {10.1016/j.nima.2018.04.009},
  url          = {https://doi.org/10.1016/j.nima.2018.04.009}
}

@article{Gurlek2025npjCompMater,
  author  = {Gurlek, Burak and Sharma, Shubham and Lazzaroni, Paolo and Rubio, Angel and Rossi, Mariana},
  title   = {Accurate machine learning interatomic potentials for polyacene molecular crystals: application to single molecule host-guest systems},
  journal = {npj Computational Materials},
  year    = {2025},
  volume  = {11},
  pages   = {318},
  doi     = {10.1038/s41524-025-01825-w},
  url     = {https://doi.org/10.1038/s41524-025-01825-w}
}

@article{Parker2017JPCC,
  author  = {Parker, Stewart F. and Parker, Jack L. and Jura, Marek},
  title   = {Structure and Vibrational Spectra of 2,5-Diiodothiophene: A Model for Polythiophene},
  journal = {The Journal of Physical Chemistry C},
  year    = {2017},
  volume  = {121},
  number  = {23},
  pages   = {12636--12642},
  doi     = {10.1021/acs.jpcc.7b03803},
  url     = {https://doi.org/10.1021/acs.jpcc.7b03803}
}

@article{Sivaraman2024JPCM,
  author  = {Sivaraman, Ganesh and Benmore, Chris J.},
  title   = {Deciphering diffuse scattering with machine learning and the equivariant foundation model: The case of molten FeO},
  journal = {Journal of Physics: Condensed Matter},
  year    = {2024},
  volume  = {36},
  number  = {15},
  pages   = {L1501},
  doi     = {10.1088/1361-648X/ad577b},
  url     = {https://doi.org/10.1088/1361-648X/ad577b}
}

@article{Ferretti2025JCIM,
  author  = {Ferretti, Alfonso and Melani, Giacomo and Benedetti, Luca and Sorodoc, Robert A. and Fortunelli, Alessandro and Brancato, Giuseppe},
  title   = {Accurate Simulations of Water and Aqueous Solutions through Fine-Tuned Dispersion-Corrected Density Functional Theory and Machine-Learning Interatomic Potentials},
  journal = {Journal of Chemical Information and Modeling},
  year    = {2025},
  volume  = {65},
  number  = {22},
  pages   = {12437--12447},
  doi     = {10.1021/acs.jcim.5c02079},
  url     = {https://doi.org/10.1021/acs.jcim.5c02079}
}

@article{Morton2025RSC_Sustainability,
  author  = {Morton, K.\ S.\ C. and O'Malley, A.\ J. and Armstrong, Jeff},
  title   = {Probing adsorption interactions of lignin derivatives in industrial zeolite catalysts through combining vibrational spectroscopy and {\em ab initio} calculations},
  journal = {RSC Sustainability},
  year    = {2025},
  volume  = {3},
  pages   = {2938--2951},
  doi     = {10.1039/D5SU00024F},
  url     = {https://doi.org/10.1039/D5SU00024F}
}

@article{Morton2025Microporous,
  author  = {Morton, K.\ S.\ C. and Appel, M. and Woodward, C.\ L.\ M. and Armstrong, J. and O'Malley, A.\ J.},
  title   = {The effect of pore structure on the local and nanoscale mobility of anisole and guaiacol in commercial zeolite catalysts},
  journal = {Microporous and Mesoporous Materials},
  year    = {2025},
  volume  = {383},
  pages   = {113388},
  doi     = {10.1016/j.micromeso.2024.113388},
  url     = {https://doi.org/10.1016/j.micromeso.2024.113388}
}

@article{Shigenobu2023JPCB,
  author  = {Shigenobu, Keisuke and Tsuzuki, Seiji and Philippi, Frederik and Sudoh, Taku and Ugata, Yosuke and Dokko, Kaoru and Watanabe, Masayoshi and Ueno, Kazuhide and Shinoda, Wataru},
  title   = {Molecular Level Origin of Ion Dynamics in Highly Concentrated Electrolytes},
  journal = {The Journal of Physical Chemistry B},
  year    = {2023},
  volume  = {127},
  number  = {48},
  pages   = {10422--10433},
  doi     = {10.1021/acs.jpcb.3c05864},
  url     = {https://doi.org/10.1021/acs.jpcb.3c05864}
}

@article{Druzbicki2021JPCL,
  author  = {Dru{\.z}bicki, Kacper and Lav{\'e}n, Rasmus and Armstrong, Jeff and Malavasi, Lorenzo and Fernandez-Alonso, Felix and Karlsson, Maths},
  title   = {Cation Dynamics and Structural Stabilization in Formamidinium Lead Iodide Perovskites},
  journal = {The Journal of Physical Chemistry Letters},
  year    = {2021},
  volume  = {12},
  number  = {22},
  pages   = {3503--3508},
  doi     = {10.1021/acs.jpclett.1c00616},
  url     = {https://doi.org/10.1021/acs.jpclett.1c00616}
}

@article{MarinVilla2025JPCL,
  author  = {Marin-Villa, Pelayo and Gila-Herranz, Pablo and Jimenez-Ruiz, Mon{\i}ca and Ivanov, Alexandre and Armstrong, Jeff and Dru{\.z}bicki, Kacper and Fernandez-Alonso, Felix},
  title   = {Molecular Derailment via Pressurization in Methylammonium Lead Iodide},
  journal = {The Journal of Physical Chemistry Letters},
  year    = {2025},
  volume  = {16},
  number  = {42},
  pages   = {10906--10914},
  doi     = {10.1021/acs.jpclett.5c01832},
  url     = {https://doi.org/10.1021/acs.jpclett.5c01832}
}

@article{Armstrong2022JPhysCommun,
  author  = {Armstrong, Jeff},
  title   = {Neutron Spectroscopy as a Method for Classical Force-Field Parameterization: Past Methods, Present Successes and Future Challenges},
  journal = {Journal of Physics Communications},
  year    = {2022},
  volume  = {6},
  number  = {10},
  pages   = {102002},
  doi     = {10.1088/2399-6528/ac9728},
  url     = {https://doi.org/10.1088/2399-6528/ac9728}
}

@article{armstrongpressure,
title = {The unlocking of high-pressure science with broadband neutron spectroscopy at the ISIS Pulsed Neutron and Muon Source},
journal = {Nuclear Instruments and Methods in Physics Research Section A: Accelerators, Spectrometers, Detectors and Associated Equipment},
volume = {1039},
pages = {167097},
year = {2022},
issn = {0168-9002},
doi = {https://doi.org/10.1016/j.nima.2022.167097},
url = {https://www.sciencedirect.com/science/article/pii/S0168900222005022},
author = {Jeff Armstrong and Xiao Wang and Felix Fernandez-Alonso}
}

@article{Mor2025Langmuir,
  author  = {Mor, Jaideep and Sharma, Sandeep K. and Bahadur, Jitendra and Kancharlapalli, Srinivasu and Armstrong, Jeff},
  title   = {Fine-Tuning of Gate Opening in Hybrid Zeolitic Imidazolate Framework-8: Inelastic Neutron Scattering and Density Functional Theory Study},
  journal = {Langmuir},
  year    = {2025},
  volume  = {41},
  number  = {39},
  pages   = {26770--26778},
  doi     = {10.1021/acs.langmuir.5c03280}
}

@article{Bahadur2025JPCL,
  author  = {Bahadur, Jitendra and Sharma, Sandeep K. and Kancharlapalli, Srinivasu and Mor, Jaideep and Armstrong, Jeff and Sen, Debasis},
  title   = {Competitive Host-Guest and Guest-Guest Interaction Dependent Flexibility of Crystal-Size-Engineered ZIF-8: An Inelastic Neutron Scattering Study},
  journal = {The Journal of Physical Chemistry Letters},
  year    = {2025},
  volume  = {16},
  number  = {22},
  pages   = {5496--5505},
  doi     = {10.1021/acs.jpclett.5c00078},
  note    = {Epub 2025-05-27}
}

@article{Schweicher2019,
  author = {
    Schweicher, Guillaume and
    D'Avino, Gabriele and
    Ruggiero, Michael T. and
    Harkin, David J. and
    Broch, Katharina and
    Venkateshvaran, Deepak and
    Liu, Guoming and
    Richard, Audrey and
    Ruzié, Christian and
    Armstrong, Jeff and
    Kennedy, Alan R. and
    Shankland, Kenneth and
    Takimiya, Kazuo and
    Geerts, Yves H. and
    Zeitler, J. Axel and
    Fratini, Simone and
    Sirringhaus, Henning
  },
  title   = {Chasing the “Killer” Phonon Mode for the Rational Design of Low‐Disorder, High‐Mobility Molecular Semiconductors},
  journal = {Advanced Materials},
  year    = {2019},
  volume  = {31},
  number  = {25},
  pages   = {1902407},
  doi     = {10.1002/adma.201902407}
}

@article{Banks2023,
  author = {Banks, Peter A. and
    D'Avino, Gabriele and
    Schweicher, Guillaume and
    Armstrong, Jeff and
    Ruzié, Christian and
    Chung, Jong-Won and
    Park, Jeong-Il and
    Sawabe, Chizuru and
    Okamoto, Toshihiro and
    Takeya, Jun and
    Sirringhaus, Henning and
    Ruggiero, Michael T.},
  title   = {Untangling the Fundamental Electronic Origins of Non-Local Electron–Phonon Coupling in Organic Semiconductors},
  journal = {Advanced Functional Materials},
  year    = {2023},
  volume  = {33},
  number  = {38},
  pages   = {2303701},
  doi     = {10.1002/adfm.202210516}
}

@article{togoFirstprinciplesPhononCalculations2023,
  title = {First-Principles {{Phonon Calculations}} with {{Phonopy}} and {{Phono3py}}},
  author = {Togo, Atsushi},
  year = {2023},
  month = jan,
  journal = {Journal of the Physical Society of Japan},
  volume = {92},
  number = {1},
  pages = {012001},
  issn = {0031-9015, 1347-4073},
  doi = {10.7566/JPSJ.92.012001},
  urldate = {2024-12-10},
  langid = {english},
}

@article{kresseEfficientIterativeSchemes1996,
  title = {Efficient Iterative Schemes for Ab Initio Total-Energy Calculations Using a Plane-Wave Basis Set},
  author = {Kresse, G. and Furthm{\"u}ller, J.},
  year = {1996},
  month = oct,
  journal = {Physical Review B},
  volume = {54},
  number = {16},
  pages = {11169--11186},
  publisher = {American Physical Society},
  doi = {10.1103/PhysRevB.54.11169},
  urldate = {2025-07-31},
}

@article{kresseUltrasoftPseudopotentialsProjector1999,
  title = {From Ultrasoft Pseudopotentials to the Projector Augmented-Wave Method},
  author = {Kresse, G. and Joubert, D.},
  year = {1999},
  month = jan,
  journal = {Physical Review B},
  volume = {59},
  number = {3},
  pages = {1758--1775},
  publisher = {American Physical Society},
  doi = {10.1103/PhysRevB.59.1758},
  urldate = {2025-07-31},
}

@article{grimmeEffectDampingFunction2011,
  title = {Effect of the Damping Function in Dispersion Corrected Density Functional Theory},
  author = {Grimme, Stefan and Ehrlich, Stephan and Goerigk, Lars},
  year = {2011},
  journal = {Journal of Computational Chemistry},
  volume = {32},
  number = {7},
  pages = {1456--1465},
  issn = {1096-987X},
  doi = {10.1002/jcc.21759},
  urldate = {2025-07-31},
  copyright = {Copyright {\copyright} 2011 Wiley Periodicals, Inc.},
  langid = {english},
}

@article{perdewGeneralizedGradientApproximation1996,
  title = {Generalized {{Gradient Approximation Made Simple}}},
  author = {Perdew, John P. and Burke, Kieron and Ernzerhof, Matthias},
  year = {1996},
  month = oct,
  journal = {Physical Review Letters},
  volume = {77},
  number = {18},
  pages = {3865--3868},
  publisher = {American Physical Society},
  doi = {10.1103/PhysRevLett.77.3865},
  urldate = {2025-07-31}
}

@article{batatiaFoundationModelAtomistic2024,
    title = {A foundation model for atomistic materials chemistry},
    journal = {The Journal of Chemical Physics},
    volume = {163},
    number = {18},
    pages = {184110},
    year = {2025},
    author = {Batatia, Ilyes and Benner, Philipp and Chiang, Yuan and Elena, Alin M. and Kovács, Dávid P. and Riebesell, Janosh and Advincula, Xavier R. and Asta, Mark and Avaylon, Matthew and Baldwin, William J. and Berger, Fabian and Bernstein, Noam and Bhowmik, Arghya and Bigi, Filippo and Blau, Samuel M. and Cărare, Vlad and Ceriotti, Michele and Chong, Sanggyu and Darby, James P. and De, Sandip and Della Pia, Flaviano and Deringer, Volker L. and Elijošius, Rokas and El-Machachi, Zakariya and Fako, Edvin and Falcioni, Fabio and Ferrari, Andrea C. and Gardner, John L. A. and Gawkowski, Mikołaj J. and Genreith-Schriever, Annalena and George, Janine and Goodall, Rhys E. A. and Grandel, Jonas and Grey, Clare P. and Grigorev, Petr and Han, Shuang and Handley, Will and Heenen, Hendrik H. and Hermansson, Kersti and Ho, Cheuk Hin and Hofmann, Stephan and Holm, Christian and Jaafar, Jad and Jakob, Konstantin S. and Jung, Hyunwook and Kapil, Venkat and Kaplan, Aaron D. and Karimitari, Nima and Kermode, James R. and Kourtis, Panagiotis and Kroupa, Namu and Kullgren, Jolla and Kuner, Matthew C. and Kuryla, Domantas and Liepuoniute, Guoda and Lin, Chen and Margraf, Johannes T. and Magdău, Ioan-Bogdan and Michaelides, Angelos and Moore, J. Harry and Naik, Aakash A. and Niblett, Samuel P. and Norwood, Sam Walton and O’Neill, Niamh and Ortner, Christoph and Persson, Kristin A. and Reuter, Karsten and Rosen, Andrew S. and Rosset, Louise A. M. and Schaaf, Lars L. and Schran, Christoph and Shi, Benjamin X. and Sivonxay, Eric and Stenczel, Tamás K. and Sutton, Christopher and Svahn, Viktor and Swinburne, Thomas D. and Tilly, Jules and van der Oord, Cas and Vargas, Santiago and Varga-Umbrich, Eszter and Vegge, Tejs and Vondrák, Martin and Wang, Yangshuai and Witt, William C. and Wolf, Thomas and Zills, Fabian and Csányi, Gábor},
    month = {11},
    issn = {0021-9606},
    doi = {10.1063/5.0297006},
    url = {https://doi.org/10.1063/5.0297006},
}

@inproceedings{batatiaMACEHigherOrder2022,
  title = {{{MACE}}: {{Higher Order Equivariant Message Passing Neural Networks}} for {{Fast}} and {{Accurate Force Fields}}},
  shorttitle = {{{MACE}}},
  booktitle = {Advances in {{Neural Information Processing Systems}}},
  author = {Batatia, Ilyes and Kovacs, David Peter and Simm, Gregor N. C. and Ortner, Christoph and Csanyi, Gabor},
  year = {2022},
  month = oct,
  urldate = {2025-07-31},
  langid = {english}
}

@article{kovacsEvaluationMACEForce2023,
  title = {Evaluation of the {{MACE}} Force Field Architecture: {{From}} Medicinal Chemistry to Materials Science},
  shorttitle = {Evaluation of the {{MACE}} Force Field Architecture},
  author = {Kov{\'a}cs, D{\'a}vid P{\'e}ter and Batatia, Ilyes and Arany, Eszter S{\'a}ra and Cs{\'a}nyi, G{\'a}bor},
  year = {2023},
  month = jul,
  journal = {The Journal of Chemical Physics},
  volume = {159},
  number = {4},
  pages = {044118},
  issn = {0021-9606},
  doi = {10.1063/5.0155322},
  urldate = {2025-07-31}
}

@article{allenOptimalDataGeneration2022,
  title = {Optimal Data Generation for Machine Learned Interatomic Potentials},
  author = {Allen, Connor and Bart{\'o}k, Albert P.},
  year = {2022},
  month = dec,
  journal = {Machine Learning: Science and Technology},
  volume = {3},
  number = {4},
  pages = {045031},
  publisher = {IOP Publishing},
  issn = {2632-2153},
  doi = {10.1088/2632-2153/ac9ae7},
  urldate = {2024-03-19},
  langid = {english},
}

@article{lloyd-williamsLatticeDynamicsElectronphonon2015,
  title = {Lattice Dynamics and Electron-Phonon Coupling Calculations Using Nondiagonal Supercells},
  author = {{Lloyd-Williams}, Jonathan H. and Monserrat, Bartomeu},
  year = {2015},
  month = nov,
  journal = {Physical Review B},
  volume = {92},
  number = {18},
  pages = {184301},
  issn = {1098-0121, 1550-235X},
  doi = {10.1103/PhysRevB.92.184301},
  urldate = {2023-06-05},
  langid = {english},
}

@article{arnoldMantidDataAnalysis2014,
    title = {Mantid—{Data} analysis and visualization package for neutron scattering and μ{SR} experiments},
    volume = {764},
    issn = {0168-9002},
    year = {2014},
    url = {https://www.sciencedirect.com/science/article/pii/S0168900214008729},
    doi = {https://doi.org/10.1016/j.nima.2014.07.029},
    journal = {Nuclear Instruments and Methods in Physics Research Section A: Accelerators, Spectrometers, Detectors and Associated Equipment},
    author = {Arnold, O. and Bilheux, J. C. and Borreguero, J. M. and Buts, A. and Campbell, S. I. and Chapon, L. and Doucet, M. and Draper, N. and Leal, R. Ferraz and Gigg, M. A. and Lynch, V. E. and Markvardsen, A. and Mikkelson, D. J. and Mikkelson, R. L. and Miller, R. and Palmen, K. and Parker, P. and Passos, G. and Perring, T. G. and Peterson, P. F. and Ren, S. and Reuter, M. A. and Savici, A. T. and Taylor, J. W. and Taylor, R. J. and Tolchenov, R. and Zhou, W. and Zikovsky, J.},
    pages = {156--166},
}

@article{hjorthlarsenAtomicSimulationEnvironment2017,
  title = {The Atomic Simulation Environment---a {{Python}} Library for Working with Atoms},
  author = {Hjorth Larsen, Ask and J{\o}rgen Mortensen, Jens and Blomqvist, Jakob and Castelli, Ivano E and Christensen, Rune and Du{\l}ak, Marcin and Friis, Jesper and Groves, Michael N and Hammer, Bj{\o}rk and Hargus, Cory and Hermes, Eric D and Jennings, Paul C and Bjerre Jensen, Peter and Kermode, James and Kitchin, John R and Leonhard Kolsbjerg, Esben and Kubal, Joseph and Kaasbjerg, Kristen and Lysgaard, Steen and Bergmann Maronsson, J{\'o}n and Maxson, Tristan and Olsen, Thomas and Pastewka, Lars and Peterson, Andrew and Rostgaard, Carsten and Schi{\o}tz, Jakob and Sch{\"u}tt, Ole and Strange, Mikkel and Thygesen, Kristian S and Vegge, Tejs and Vilhelmsen, Lasse and Walter, Michael and Zeng, Zhenhua and Jacobsen, Karsten W},
  year = 2017,
  month = jul,
  journal = {Journal of Physics: Condensed Matter},
  volume = {29},
  number = {27},
  pages = {273002},
  issn = {0953-8984, 1361-648X},
  doi = {10.1088/1361-648X/aa680e},
  urldate = {2022-11-14},
}

@article{scikit-learn,
    title={Scikit-learn: Machine Learning in {P}ython},
    author={Pedregosa, F. and Varoquaux, G. and Gramfort, A. and Michel, V.
            and Thirion, B. and Grisel, O. and Blondel, M. and Prettenhofer, P.
            and Weiss, R. and Dubourg, V. and Vanderplas, J. and Passos, A. and
            Cournapeau, D. and Brucher, M. and Perrot, M. and Duchesnay, E.},
    journal={Journal of Machine Learning Research},
    volume={12},
    pages={2825--2830},
    year={2011}
   }

@article{caldeweyherGenerallyApplicableAtomiccharge2019,
  title = {A Generally Applicable Atomic-Charge Dependent {{London}} Dispersion Correction},
  author = {Caldeweyher, Eike and Ehlert, Sebastian and Hansen, Andreas and Neugebauer, Hagen and Spicher, Sebastian and Bannwarth, Christoph and Grimme, Stefan},
  year = 2019,
  month = apr,
  journal = {The Journal of Chemical Physics},
  volume = {150},
  number = {15},
  pages = {154122},
  issn = {0021-9606},
  doi = {10.1063/1.5090222},
  urldate = {2026-03-04}
}

@article{ehlertR2SCAND4DispersionCorrected2021,
  title = {{{r2SCAN-D4}}: {{Dispersion}} Corrected Meta-Generalized Gradient Approximation for General Chemical Applications},
  shorttitle = {{{r2SCAN-D4}}},
  author = {Ehlert, Sebastian and Huniar, Uwe and Ning, Jinliang and Furness, James W. and Sun, Jianwei and Kaplan, Aaron D. and Perdew, John P. and Brandenburg, Jan Gerit},
  year = 2021,
  month = feb,
  journal = {The Journal of Chemical Physics},
  volume = {154},
  number = {6},
  pages = {061101},
  issn = {0021-9606},
  doi = {10.1063/5.0041008},
  urldate = {2026-03-04}
}

@article{furnessAccurateNumericallyEfficient2020,
  title = {Accurate and {{Numerically Efficient r2SCAN Meta-Generalized Gradient Approximation}}},
  author = {Furness, James W. and Kaplan, Aaron D. and Ning, Jinliang and Perdew, John P. and Sun, Jianwei},
  year = 2020,
  month = oct,
  journal = {The Journal of Physical Chemistry Letters},
  volume = {11},
  number = {19},
  pages = {8208--8215},
  publisher = {American Chemical Society},
  doi = {10.1021/acs.jpclett.0c02405},
  urldate = {2026-03-04}
}

@misc{batatia2025crosslearningelectronicstructure,
      title={Cross Learning between Electronic Structure Theories for Unifying Molecular, Surface, and Inorganic Crystal Foundation Force Fields}, 
      author={Ilyes Batatia and Chen Lin and Joseph Hart and Elliott Kasoar and Alin M. Elena and Sam Walton Norwood and Thomas Wolf and Gábor Csányi},
      year={2025},
      eprint={2510.25380},
      archivePrefix={arXiv},
      primaryClass={physics.chem-ph},
      url={https://arxiv.org/abs/2510.25380}, 
}

@misc{batatia2026macepolar1polarisableelectrostaticfoundation,
      title={MACE-POLAR-1: A Polarisable Electrostatic Foundation Model for Molecular Chemistry}, 
      author={Ilyes Batatia and William J. Baldwin and Domantas Kuryla and Joseph Hart and Elliott Kasoar and Alin M. Elena and Harry Moore and Mikołaj J. Gawkowski and Benjamin X. Shi and Venkat Kapil and Panagiotis Kourtis and Ioan-Bogdan Magdău and Gábor Csányi},
      year={2026},
      eprint={2602.19411},
      archivePrefix={arXiv},
      primaryClass={physics.chem-ph},
      url={https://arxiv.org/abs/2602.19411}, 
}

@article{grimme2010,
    author = {Grimme, Stefan and Antony, Jens and Ehrlich, Stephan and Krieg, Helge},
    title = {A consistent and accurate ab initio parametrization of density functional dispersion correction (DFT-D) for the 94 elements H-Pu},
    journal = {The Journal of Chemical Physics},
    volume = {132},
    number = {15},
    pages = {154104},
    year = {2010},
    month = {04},
    issn = {0021-9606},
    doi = {10.1063/1.3382344},
    url = {https://doi.org/10.1063/1.3382344},
    eprint = {https://pubs.aip.org/aip/jcp/article-pdf/doi/10.1063/1.3382344/15684000/154104_1_online.pdf},
}

@article{Larsen2017,
doi = {10.1088/1361-648X/aa680e},
url = {https://doi.org/10.1088/1361-648X/aa680e},
year = {2017},
month = {jun},
publisher = {IOP Publishing},
volume = {29},
number = {27},
pages = {273002},
author = {Hjorth Larsen, Ask and Jørgen Mortensen, Jens and Blomqvist, Jakob and Castelli, Ivano E and Christensen, Rune and Dułak, Marcin and Friis, Jesper and Groves, Michael N and Hammer, Bjørk and Hargus, Cory and Hermes, Eric D and Jennings, Paul C and Bjerre Jensen, Peter and Kermode, James and Kitchin, John R and Leonhard Kolsbjerg, Esben and Kubal, Joseph and Kaasbjerg, Kristen and Lysgaard, Steen and Bergmann Maronsson, Jón and Maxson, Tristan and Olsen, Thomas and Pastewka, Lars and Peterson, Andrew and Rostgaard, Carsten and Schiøtz, Jakob and Schütt, Ole and Strange, Mikkel and Thygesen, Kristian S and Vegge, Tejs and Vilhelmsen, Lasse and Walter, Michael and Zeng, Zhenhua and Jacobsen, Karsten W},
title = {The atomic simulation environment—a Python library for working with atoms},
journal = {Journal of Physics: Condensed Matter}
}

@software{kasoar_2026_18713710,
  author       = {Kasoar, Elliott and
                  Austin, Patrick and
                  Devereux, Harvey and
                  Harris, Kieran and
                  Mason, David and
                  Wilkins, Jacob and
                  Zanca, Federica and
                  Elena, Alin},
  title        = {janus-core},
  year         = 2024,
  publisher    = {Zenodo},
  version      = {v0.8.7},
  doi          = {10.5281/zenodo.18713710},
  url          = {https://doi.org/10.5281/zenodo.18713710},
}

@misc{kaplan2025foundationalpotentialenergysurface,
      title={A Foundational Potential Energy Surface Dataset for Materials}, 
      author={Aaron D. Kaplan and Runze Liu and Ji Qi and Tsz Wai Ko and Bowen Deng and Janosh Riebesell and Gerbrand Ceder and Kristin A. Persson and Shyue Ping Ong},
      year={2025},
      eprint={2503.04070},
      archivePrefix={arXiv},
      primaryClass={cond-mat.mtrl-sci},
      url={https://arxiv.org/abs/2503.04070}, 
}

@misc{levine2025openmolecules2025omol25,
      title={The Open Molecules 2025 (OMol25) Dataset, Evaluations, and Models}, 
      author={Daniel S. Levine and Muhammed Shuaibi and Evan Walter Clark Spotte-Smith and Michael G. Taylor and Muhammad R. Hasyim and Kyle Michel and Ilyes Batatia and Gábor Csányi and Misko Dzamba and Peter Eastman and Nathan C. Frey and Xiang Fu and Vahe Gharakhanyan and Aditi S. Krishnapriyan and Joshua A. Rackers and Sanjeev Raja and Ammar Rizvi and Andrew S. Rosen and Zachary Ulissi and Santiago Vargas and C. Lawrence Zitnick and Samuel M. Blau and Brandon M. Wood},
      year={2025},
      eprint={2505.08762},
      archivePrefix={arXiv},
      primaryClass={physics.chem-ph},
      url={https://arxiv.org/abs/2505.08762}, 
}

@Article{LAMMPS,
  author = "A. P. Thompson and H. M. Aktulga and R. Berger and 
     D. S. Bolintineanu and W. M. Brown and P. S. Crozier and
     P. J. in 't Veld and A. Kohlmeyer and S. G. Moore and T. D. Nguyen and
     R. Shan and M. J. Stevens and J. Tranchida and C. Trott and S. J. Plimpton",
  title = "{LAMMPS} - a flexible simulation tool for
     particle-based materials modeling at the 
     atomic, meso, and continuum scales",
  journal = "Comp. Phys. Comm.",
  volume =  "271",
  pages =   "108171",
  year =    "2022",
  doi = "10.1016/j.cpc.2021.108171"
}

\end{document}

\maketitle

\begingroup
\renewcommand\thefootnote{}
\footnotetext{Jeff Armstrong ORCID: 0000-0002-8326-3097}
\footnotetext{Adam Jackson ORCID: 0000-0001-5272-6530}
\footnotetext{Alin Elena ORCID: 0000-0002-7013-6670}
\endgroup

\vspace{-4em}
\tableofcontents

\newpage

% ============================================================
\section{Experimental methods}

\subsection{Sample and measurement conditions}
All \gls{ins} measurements were performed on the TOSCA inelastic neutron spectrometer at the ISIS Neutron and Muon Source.\cite{Pinna2018NeutronGuideTOSCA}
Ambient-pressure spectra were collected at \SI{10}{\kelvin} using a \SI{2}{\gram} sample of crystalline 2,5-diiodothiophene loaded into a standard aluminium sample can.
High-pressure spectra were collected at \SI{10}{\kelvin} at \SI{1.5}{\giga\pascal} using a \SI{200}{\milli\gram} sample loaded into a low-background NiCrAl clamp cell.
The use of cryogenic temperature suppresses large-amplitude molecular motion, thus minimizing the Debye-Waller peak suppression and improving the quantitative comparison with harmonic lattice-dynamical calculations.

\subsection{High-pressure clamp cell and masking}
The \SI{1.5}{\giga\pascal} measurements were carried out in a NiCrAl alloy clamp cell designed to minimise parasitic neutron scattering while retaining mechanical robustness at gigapascal pressures.\cite{armstrongpressure}
To further suppress scattering from the bulk body of the clamp assembly, a cadmium mask was employed to restrict the illuminated volume to the sample region.
In practice, the residual cell background depends weakly on small differences in mask positioning, sample centreing, and multiple scattering contributions.

\subsection{Pressure generation and reproducibility}
Pressure was applied ex situ by mechanical loading of the clamp cell with 3~tonnes of weights.
The reported pressure corresponds to the nominal applied load for the calibrated clamp geometry when cooled to 10~K.

% ============================================================
\section{Data reduction and background subtraction}

\subsection{Empty-cell subtraction}
For the high-pressure measurement, cell-background subtraction was performed using an independently measured empty-cell spectrum acquired under nominally identical instrument settings.
Because the \SI{1.5}{\giga\pascal} dataset has a substantially lower sample mass and additional attenuation through the pressure cell, the high-pressure spectra exhibit an elevated statistical noise floor relative to the ambient-pressure dataset. Nevertheless, peak positions and relative pressure-dependent shifts remain well-resolved across the \SIrange{0}{1200}{\per\centi\meter} range.

\subsection{Practical considerations and uncertainty}
Residual artefacts after subtraction can arise from:
\begin{itemize}
  \item small differences in the effective illuminated volume (mask and cell positioning),
  \item multiple scattering contributions that depend on sample loading,
\end{itemize}

\begin{figure}[htbp]
 \centering
 \includegraphics[width=0.65\linewidth]{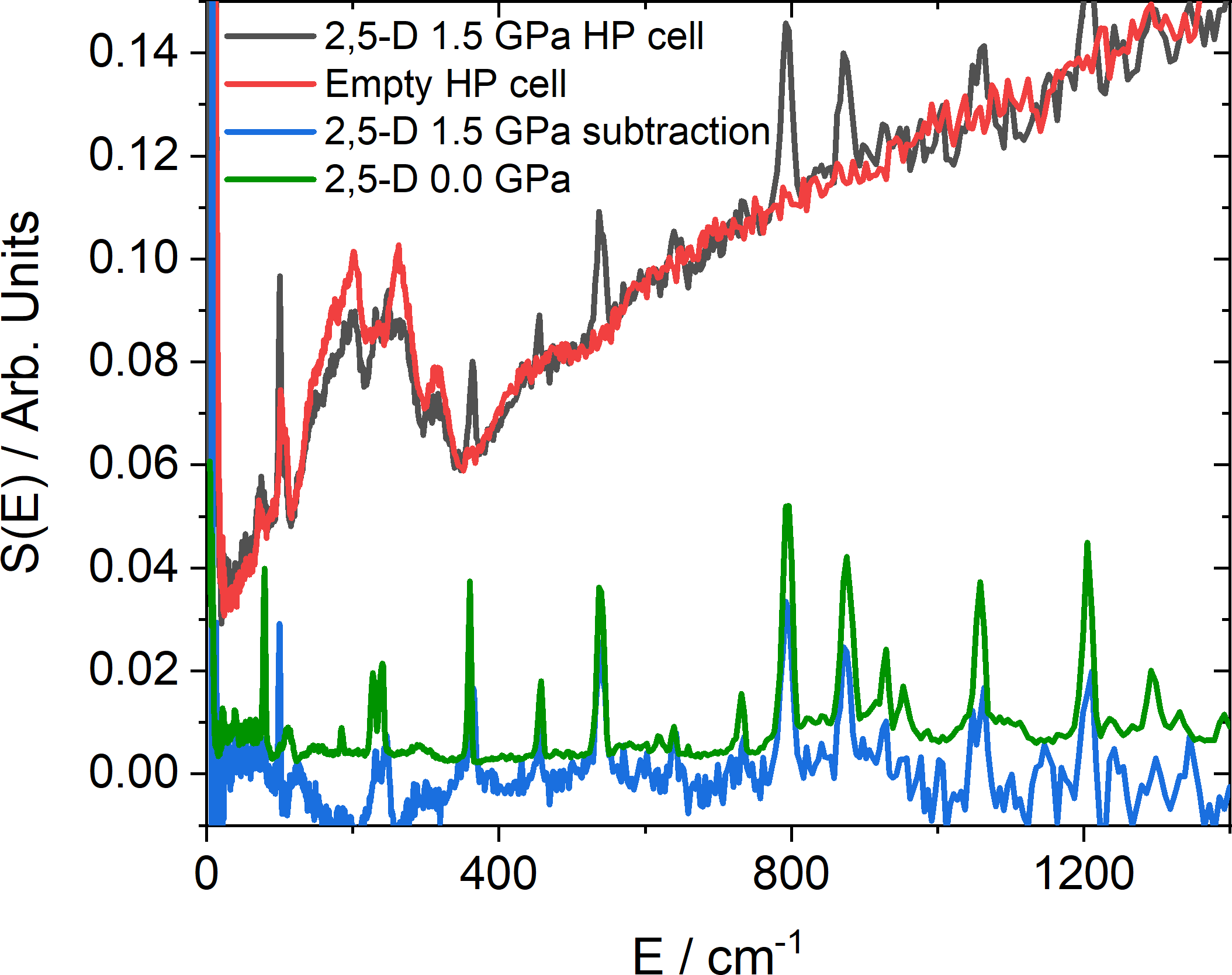}
 \caption{Subtracted TOSCA INS spectra of 2,5-D at ambient pressure (green) is compared to the 1.5~GPa measurement (blue). Unsubtracted data at 1.5~GPa (black) and the accompanying empty pressure cell (red) are included to give the reader insight into the background.}
 \label{exptsub}
\end{figure}

\begin{figure}[htbp]
 \centering
 \includegraphics[width=0.65\linewidth]{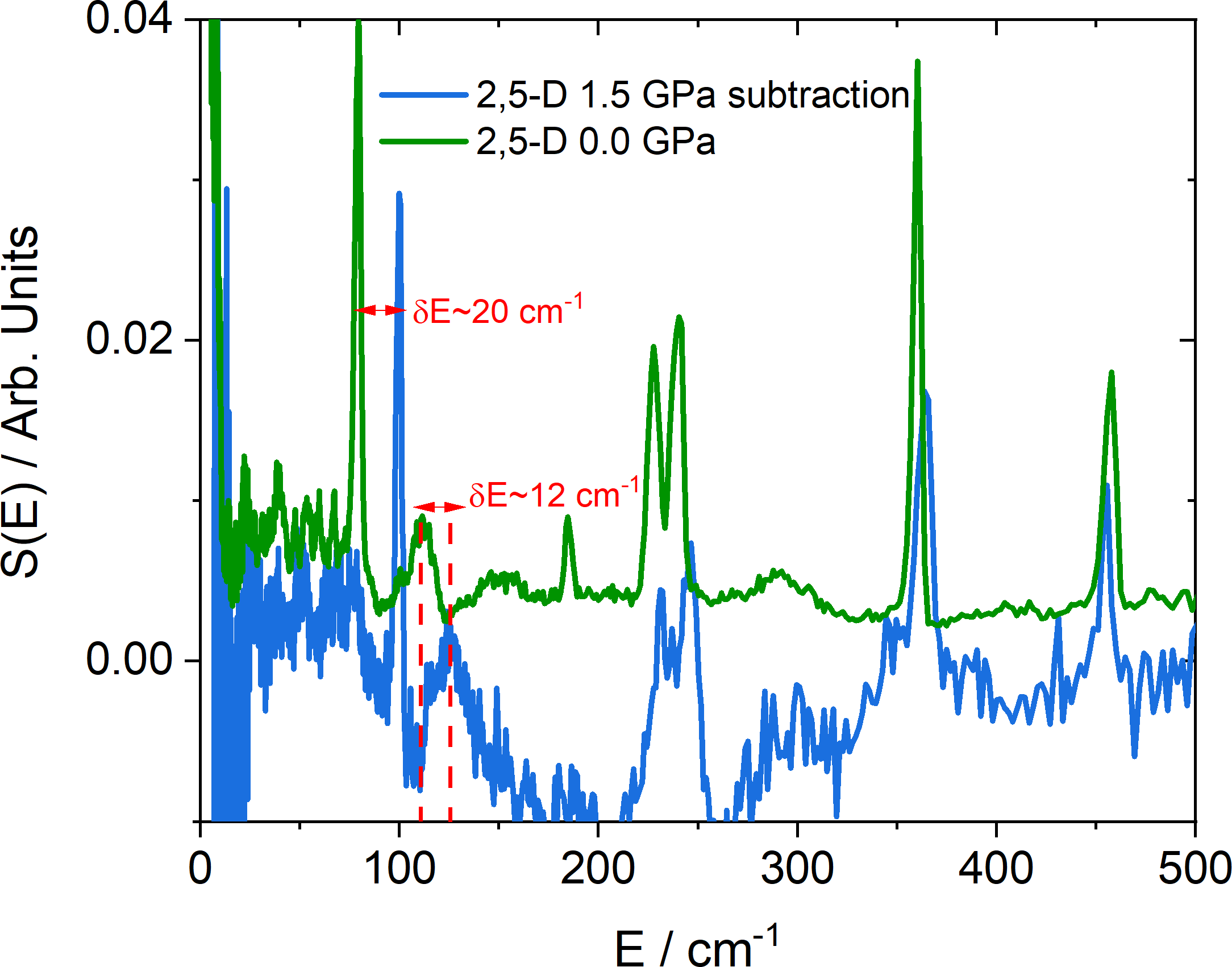}
 \caption{An energy expanded plot of the two fully subtracted spectra; ambient pressure in green, 1.5~GPa in blue.}
 \label{exptlow}
\end{figure}

Figure~\ref{exptsub} shows a comparison of the cell signal and sample signals before and after subtraction, to give the reader an idea of the relative intensities. A zoomed in plot of the experimental subtracted spectra is shown in Figure~\ref{exptlow}.

% ============================================================
\section{Density-functional theory calculations}

\subsection{Electronic structure parameters}
\Gls{dft} calculations were performed using the \gls{paw} method as implemented in \vasp{}.\cite{kresseEfficientIterativeSchemes1996,kresseUltrasoftPseudopotentialsProjector1999}
Structures were optimised at the target pressures by applying an external stress and relaxing both lattice parameters and atomic positions.

A plane-wave cutoff energy of \SI{520}{\eV} was used for ``PBE-D3'' calculations with the PBE \gls{xc} functional and Grimme D3 correction including Becke--Johnson damping.\cite{perdewGeneralizedGradientApproximation1996,grimmeEffectDampingFunction2011}.
Reciprocal space was sampled automatically with a target spacing of \SI{0.2}{\per\angstrom}, shifted away from the \gammapoint{}.
Electronic self-consistency was converged to \SI{1e-7}{\eV} per calculation cell or less, using Gaussian smearing of \SI{0.01}{\eV}.
The cutoff energy and \gls{paw} datasets were selected to remain consistent with the reference setup used for the \macemp{} foundation model training data, ensuring compatibility between the baseline electronic-structure description and the MLIP fine-tuning workflow.\cite{batatiaFoundationModelAtomistic2024}

For the R2SCAN-D4 dataset slightly different parameters were used
(including a \SI{680}{\eV} plane-wave cutoff and $\mathbf{k}$-point sampling offset to $\Gamma$)
as documented for the MatPES dataset, \cite{kaplan2025foundationalpotentialenergysurface}
to which a compatible D4 correction was added.\cite{ehlertR2SCAND4DispersionCorrected2021}

\subsection{Functional sensitivity}
\begin{figure}[htbp]
 \centering
 \includegraphics[width=0.85\linewidth]{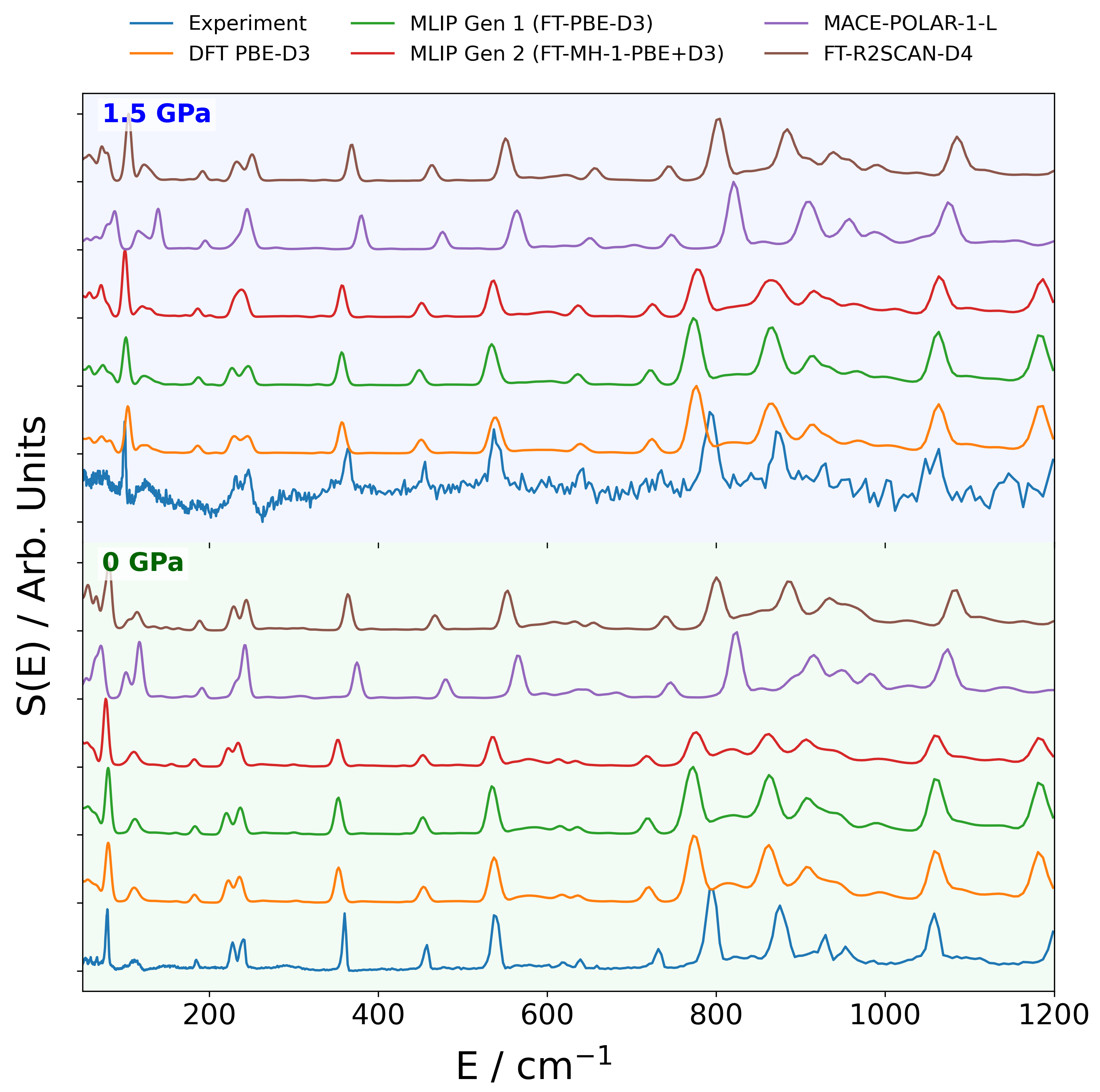}
 \caption{Vibrational spectra $S(E)$ for experiment, DFT, finetuned MLIPs and newly released polarisable molecular foundation model MACE-POLAR-1-L. Data for the two pressures 0~GPa and 1.5~GPa are shown in the lower green panel and upper blue panel respectively.}
 \label{SIplots}
\end{figure}

To assess sensitivity to the underlying electronic-structure description, additional models were fine tuned at different levels of theory and fine tuning methodology, see details in the next section of the models. Calculations were performed using the FT-MH-1-PBE-D3, MACE-POLAR-1-L, and FT-R2SCAN-D4, while keeping all other numerical settings fixed.
This provides a controlled basis for interpreting differences between refined \mlips{} as differences in the reference potential-energy surface, rather than differences in convergence settings. Figure~\ref{SIplots} shows the various MLIP spectra from our study. We see that the fine tuned model R2SCAN-D4 also produces very similar spectra to that of the PBE-D3(BJ) used in the main text, while the foundation model MACE-POLAR-1-L despite including electrostatics and using OMOL25 dataset\cite{levine2025openmolecules2025omol25} at $\omega$B97M-V/def2-TZVPD level of theory is not competitive against short range fine tuned models without further system specific fine tuning. 

\section{MLIP training protocol and dataset construction}

\subsection{Overview}
MLIP calculations were performed using the \mace{} architecture with GPU acceleration via PyTorch.\cite{batatiaMACEHigherOrder2022,kovacsEvaluationMACEForce2023}
A system-specific potential was obtained by fine-tuning the MACE-MP-0b3 ``foundation model'' on targeted \gls{dft} data.\cite{batatiaFoundationModelAtomistic2024}
The guiding objective of the refinement protocol is twofold: (i) reproduce the harmonic vibrational response at both zero pressure and \SI{1.5}{\giga\pascal}, and (ii) ensure stability and physically sensible behaviour under finite-temperature sampling (S7).

\subsection{Training configurations and fine-tuning}

Training configurations were generated via four techniques. 

 \textbf{Geometry optimisation} of the the perfect crystal using DFT at PBE-D3 level, then from the resulting trajectory only frames that differ by more than 50 meV were kept in the training set.
 
\textbf{Volume Scans}, using equation of state approach we compressed the initial crystal by 10\%  of its volume and relaxed to the same amount in 11 steps. For all these single point calculations were performed and data added in the training set. 

\textbf{Molecular dynamics:} NPT trajectories were run under a Nose--Hoover thermostat at 300~K on a unit-cell using janus\_core, with a length of max 1 ns saving configurations with a 1 ps frequency, using different pressures, 0, 0.5, 1.0 and 1.5, 2.0 GPa. Each such generated trajectory has MACE descriptors computed on each frame and further point sampling is used to sample a number of frames (max 100) in this reduced space. The selected frames are split in and 8:1:1 ratio between training, validation and testing sets.

%Initial training configurations comprised 288-atom supercells generated from 
%$4\times1\times1$ expansions of the crystallographic unit cell with controlled scaling and skewing to sample local strain environments.
%To move beyond purely harmonic neighbourhood sampling, refinement was performed iteratively and supplemented by additional configurations designed to span:
%\begin{itemize}
%  \item anisotropic strain states relevant to pressure-driven lattice contraction,
%  \item local distortions that probe coupling between intra- and intermolecular degrees of freedom,
%  \item finite-temperature fluctuations generated from short NPT \gls{md} trajectories at \SI{300}{\kelvin}.
%\end{itemize}

\textbf{Non-diagonal supercells:}
After five initial training cycles, additional 288-atom configurations were generated using alternative supercell matrices originally developed for efficient phonon sampling and recently repurposed for MLIP training.\cite{lloyd-williamsLatticeDynamicsElectronphonon2015,allenOptimalDataGeneration2022}
The supercell matrices
\[
\left( \begin{array}{ccc}
2 & 1 & 0 \\
2 & -1 & 0 \\
0 & 0 & 1
\end{array} \right),\quad
\left( \begin{array}{ccc}
4 & 0 & 0 \\
0 & 1 & 0 \\
2 & 0 & -1
\end{array} \right),\quad
\left( \begin{array}{ccc}
2 & -1 & 0 \\
-2 & 0 & 1 \\
2 & 1 & 0
\end{array} \right)
\]
were used to construct simulation cells.
Molecular dynamics was run in the NPT ensemble using ASE at \SI{300}{\kelvin} and \SI{0}{\giga\pascal} for \num{20000} steps with a \SI{1}{\femto\second} time step.
Representative snapshots were labelled with \gls{dft} energies, forces, and stresses, and incorporated into subsequent training iterations.
These samples were intended to identify/address any finite-size effects related to the consistent use of particular periodic boundaries. At the end of the process, we have 1412 structures in the training set, 176 in test and 176 in validation set. 

The first fine-tuned model involved naive fine-tuning against mace\_mp\_0b3, using a loss stress and learning rate of 0.005 and an ema\_decay of 0.995,  a batch size of 8 and 2000 epochs. Fittings involved an energy RMSE of \SI{10.0}{\meV \per \atom}, \SI{8.4}{\meV\angstrom^{-1}} for forces, and \SI{0.7}{\meV\angstrom^{-3}} for stresses.
We used an A100 NVIDIA GPU for the fine tuning with a runtime of 20s per epoch. While this was the preferred method at the time of finetuning, naive fine-tuning is associated with the risk of ``catastrophic forgetting''\cite{batatia2025crosslearningelectronicstructure}.
This is the first generation model in the main text (i.e. \textbf{FT-PBE-D3}). No observed issues were assigned to catastrophic forgetting during this work.

For fine tuning against R2SCAN-D4 DFT we selected randomly 1068 of the existing structures for training and 133 for the validation and test sets, and recomputed the energy, forces and stress at the new level of theory. We performed multihead fine-tuning against mace\_omat-0, using loss universal and learning rate of 0.0001 and an ema\_decay of 0.99999, a batch size of 2 and 10 epochs.
The replay dataset used was a combination of Materials project and OMAT datasets, as available on MACE website, and we used the original DFT labels.  The fitting results were \SI{1.05}{\meV\per\atom} RMSE of energy, \SI{26.59}{\meV\angstrom^{-1}} RMSE of forces, and \SI{0.48}{\meV\angstrom^{-3}} RMSE of stresses.
We used 4 A100 NVIDIA GPUs for the fine tuning with a runtime of 1h and 18 minutes per epoch. The resulting model is called \textbf{FT-R2SCAN-D4}.

The third finetuning strategy used the same structures as in R2SCAN-D4 case but at PBE-D3 level of theory.  We performed multihead fine-tuning against mh-1 model, head omat\_pbe, using loss stress and learning rate of 0.0001 and an ema\_decay of 0.99999, a batch size of 2 and 10 epochs, for second stage fine tuning use reduced learning rate by one order of magnitude. The replay dataset used Materials project dataset, from which we selected 30000 samples that contain combinations of elements with atomic numbers 1, 6, 16, 53, and we used pseudolabels computed with mh-1 model, head omat\_pbe for these structures. Fitting results were RMSE for eneregy 0.94 meV/Atom, for forces an RMSE of 22.24 meV / \AA, and for stress an of 0.30 meV / \AA$^3$.  We used a 4xA100 NVIDIA GPU for the fine tuning with a runtime of 56 minutes per epoch. This is the model to produce all the data for molecular dynamics simulations in the main paper. We will call this model \textbf{FT-MH-1-PBE-D3}, and it is also compared in Figure~1 of the main text to the experimental phonons.

\subsection{Practical notes on transferability}
For molecular crystals, the most demanding aspects of the potential-energy landscape typically involve weak intermolecular dispersion, steric contacts, and subtle pressure-driven reorientation within the packing motif.
The refinement protocol therefore deliberately includes configurations that perturb intermolecular separations and orientations, which are the primary drivers of the pressure derivatives observed in the low-energy vibrational region.

\section{Structural data comparison}

\begin{table}[h]
    \centering
    \caption{Lattice parameters at different pressures and simulation methods.}
    
    \vspace{0.5em}
    \textbf{Pressure: 0 GPa} \\
    \begin{tabular}{lcccccc}
        \toprule
        Method & $a$ [\AA] & $b$ [\AA] & $c$ [\AA] & $\alpha$ [$^\circ$] & $\beta$ [$^\circ$] & $\gamma$ [$^\circ$] \\
        \midrule
        exp & 14.9443(9) & 18.2941(15) & 5.2541(3) & 90.0 & 90.0 & 90.0 \\
        DFT (PBE-D3) & 14.895 & 18.123 & 5.136 & 90.0 & 90.0 & 90.0 \\
        MACE\_MP-0b3-D3* & 14.694 & 17.988 & 5.166 & 90.0 & 90.0 & 90.0 \\
        MACE\_MP-0b3-D3 & 12.705 & 18.236 & 5.702 & 90.0 & 90.0 & 90.0 \\
        MACE\_MP-OMAT-0-D3 &  15.776  &  17.199       &   5.543    & 90.0 & 90.0  & 90.0 \\
        MACE-MH-1 OMAT\_PBE-D3 & 13.192 & 18.141 & 5.825& 90.0 &90.0   &90.0 \\
        MACE-POLAR-1-L &  12.729 & 18.240& 5.921 & 90.0 & 90.0 & 90.0\\
        FT-PBE-D3 & 14.915 & 18.114 & 5.129 & 90.0 & 90.0 & 90.0 \\
        FT-R2SCAN-D4 & 14.745   &   18.382      &  5.301     & 90.0 & 90.0 & 90.0 \\
        FT-MH-1-PBE-D3 & 14.606   &    18.165     &  5.251     & 90.0 & 90.0 & 90.0\\
        FT-MH-1-PBE-D3* & 14.791   &    18.107     &  5.200     & 90.0 & 90.0 & 90.0\\
        \bottomrule
    \end{tabular}

    \vspace{1.5em}
    
    \textbf{Pressure: 1.5 GPa} \\
    \begin{tabular}{lcccccc}
        \toprule
        Method & $a$ [\AA] & $b$ [\AA] & $c$ [\AA] & $\alpha$ [$^\circ$] & $\beta$ [$^\circ$] & $\gamma$ [$^\circ$] \\
        \midrule
        exp & --- & --- & --- & --- & --- & --- \\
        DFT (PBE-D3) & 14.310 & 17.322 & 5.008 & 90.0 & 90.0 & 90.0 \\
        MACE\_MP-0b3-D3* & 14.088 & 17.246 & 4.953 & 90.0 & 90.0 & 90.0 \\
        MACE\_MP-0b3-D3 & 11.872 & 17.329 & 5.723 & 90.0 & 90.0 & 90.0 \\
        MACE\_MP-OMAT-0-D3 & 13.454   &    17.266     &   5.580    & 90.0 & 90.0 & 90.0 \\
        MACE-MH-1 OMAT\_PBE-D3 & 12.817 & 17.311 & 5.607 & 90.0 & 90.0 & 90.0 \\
        MACE-POLAR-1-L & 12.385 & 17.494 &  5.762 & 90.0 & 90.0 & 90.0\\
        FT-PBE-D3 & 14.358 & 17.335 & 4.989 & 90.0 & 90.0 & 90.0 \\
        FT-R2SCAN-D4 & 14.121   &  17.499       &   5.189    & 90.0 & 90.0 & 90.0\\
        FT-MH-1-PBE-D3 & 14.060   &  17.348       &    5.112   &  90.0 &90.0  &90.0 \\
        FT-MH-1-PBE-D3* & 14.265   &  17.463       &    5.015   &  90.0 &90.0  &90.0 \\
        \bottomrule
    \end{tabular}
    
    \vspace{0.5em}
    \small
    *Only cell vectors and positions were optimised; angles were fixed.* 
    \end{table}

% ============================================================
\section{Harmonic lattice dynamics and mode assignment}

Phonon force constants were obtained from MLIPs using finite displacements with \phonopy{}.\cite{togoFirstprinciplesPhononCalculations2023}
Peak assignments were obtained from the $\Gamma$-point MLIP phonon eigenvectors at \SI{0}{\giga\pascal}, guided by atomic displacement character and visual inspection of eigenmodes.
The dominant pressure response below \SI{500}{\per\centi\meter} arises from modes that modulate intermolecular contacts and steric repulsion within the herringbone-like packing motif, while the anomalous red shift near \SI{453}{\per\centi\meter} corresponds to an out-of-plane deformation of the thiophene ring (C--S torsional character).

\begin{table}[!htbp]
\centering
\caption{Peak assignments from \SI{0}{\giga\pascal} MLIP phonon calculations.}
\label{tab:vibrations}
\begin{tabular}{ll}
\toprule
Peak position / cm$^{-1}$ & Description \\
\midrule
80 & Libration \\
105--116 & C--I in-phase out-of-plane and in-plane bend \\
182 & C--I symmetric stretch \\
220 & C--I out-of-phase out-of-plane bend \\
237 & C--I out-of-phase in-plane bend \\
353 & C--I antisymmetric stretch \\
453 & Out-of-plane ring deformation (C--S torsion) \\
\bottomrule
\end{tabular}
\end{table}

\begin{figure}[htbp]
 \centering \includegraphics[width=0.8\linewidth]{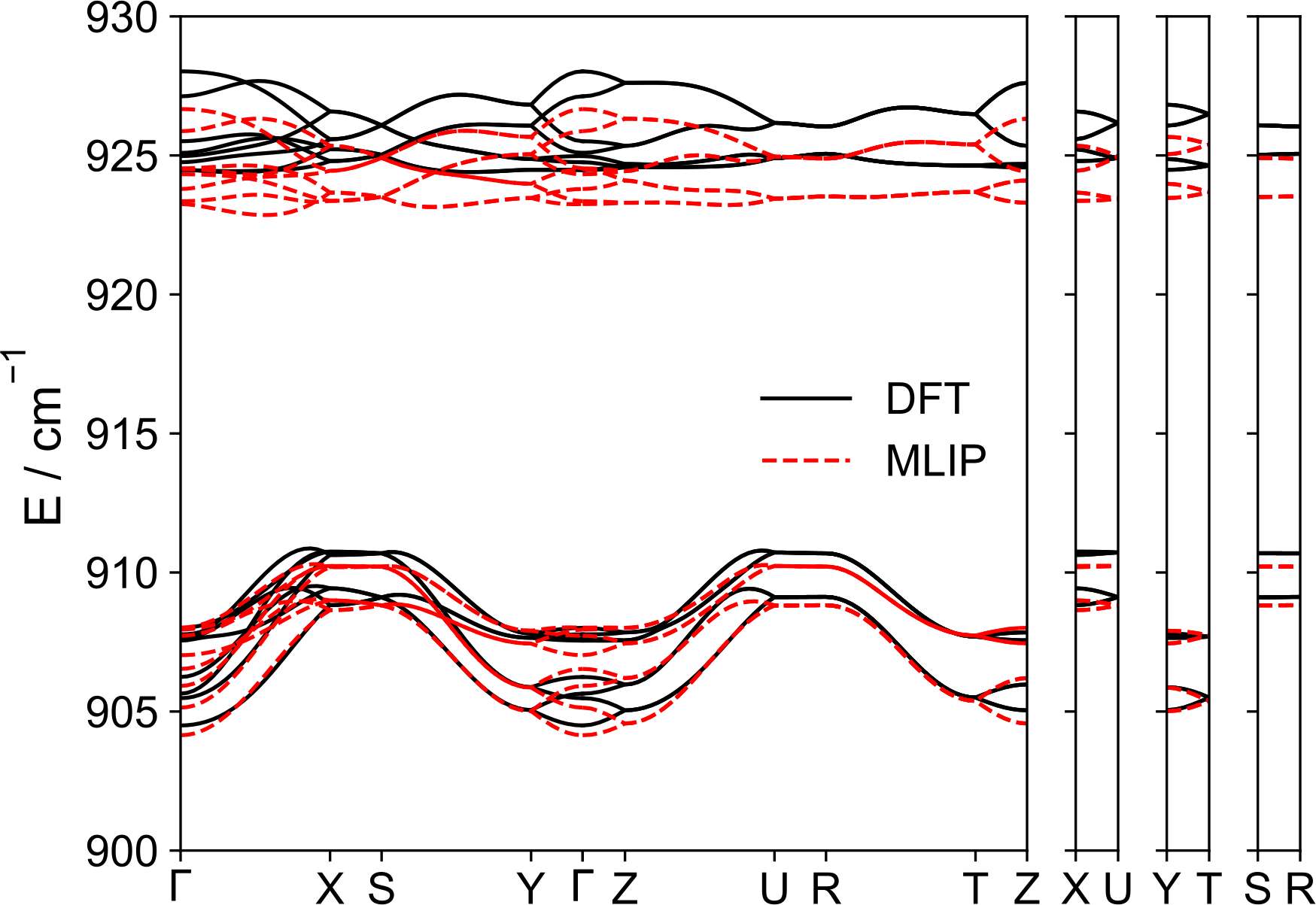}
 \caption{A direct comparison of the DFT and MLIP phonon bands. We have zoomed into the deformation region to allow the comparison of the subtle dispersion pattern in this region.}
 \label{bands}
\end{figure}

In addition to gamma point phonon calculations, we also calculated phonon dispersion curves to compare with DFT. This provides a particularly fine test of the reproduction of the dynamics, as the dispersion patterns are particularly fine features. Figure~\ref{bands} shows the band diagram of this comparison.

% ============================================================
\section{INS simulations with \abins{}}

INS spectra from simulated data were produced using the \abins{} code distributed as part of \mantid{}\cite{Dymkowski2018AbINS,arnoldMantidDataAnalysis2014}.
Phonon force constants were obtained from the MLIPs using \phonopy{} via the \codename{janus-core} command-line tools.
Force constants were converted into phonon frequencies and eigenvectors, which were then used by \abins{} to compute neutron-weighted vibrational spectra appropriate for direct comparison with TOSCA measurements.
\abins{} employs a \gls{dos}-like incoherent approximation to the INS spectrum, computing intensities along the kinematically constrained $\omega$--$Q$ trajectories of each TOSCA detector bank, and applies an energy-dependent Gaussian resolution function.
The one-phonon spectrum and second-order combination modes were enumerated within the ``almost-isotropic approximation'', and a fully isotropic convolution approximation was used to include higher quantum orders (3--10). These higher-order contributions are important for reproducing the elevated background at higher energies and for ensuring consistent intensity scaling across the full \SIrange{0}{1200}{\per\centi\meter} range.

% ============================================================
\section{Finite-temperature molecular dynamics validation}

\subsection{Simulation protocol}
To assess thermodynamic stability beyond the harmonic limit, the final MLIP was validated using \gls{md} simulations at \SI{300}{\kelvin} under periodic boundary conditions, employing a 1152-atom supercell.
The system was equilibrated in the NPT ensemble for 200~ps, followed by a production run in the NVT ensemble of 1~ns to collect statistics.
Stability was assessed through complementary structural, thermodynamic, and dynamical metrics, including the \gls{msd}, stress tensor fluctuations, potential energy stationarity, and \gls{rdf} invariance over time.

\subsection{Metrics and interpretation}
The integrity of the molecular geometry and packing motif was monitored using intra- and intermolecular \glspl{rdf} $g(r)$ for C--C, C--H, and H--H pairs.
To explicitly test for temporal drift, $g(r)$ was evaluated in an early (0.1--0.3~ns) and late (0.8--1.0~ns) window of the NVT trajectory; indistinguishable distributions indicate preservation of bonding environments and intermolecular packing.
Stress tensor components exhibit stationary fluctuations without systematic drift, consistent with mechanical stability of the condensed phase.
Potential energy fluctuations remain bounded around a well-defined mean, indicating absence of slow relaxation or instability.
Finally, \glspl{msd} for all atomic species plateau at physically reasonable values, consistent with bounded vibrational and librational motion and the absence of diffusive runaway behaviour under these conditions.

\begin{table}[!htbp]
\centering
\caption{Summary of finite-temperature \gls{md} validation conditions (reported as used in the main manuscript).}
\label{tab:md_settings}
\begin{tabular}{ll}
\toprule
Quantity & Value \\
\midrule
Supercell size & 1152 atoms \\
Temperature & \SI{300}{\kelvin} \\
Equilibration & NPT, \SI{500}{\pico\second} (main text) \\
Production & NVT, \SI{1}{\nano\second} (main text) \\
Time step & \SI{1}{\femto\second} \\
Key observables & \gls{msd}, stress tensor, $\Delta U$, \gls{rdf} invariance \\
\bottomrule
\end{tabular}
\end{table}

\section{Equation-of-state analysis}
\begin{figure}[htbp]
 \centering
 \includegraphics[width=0.4\linewidth]{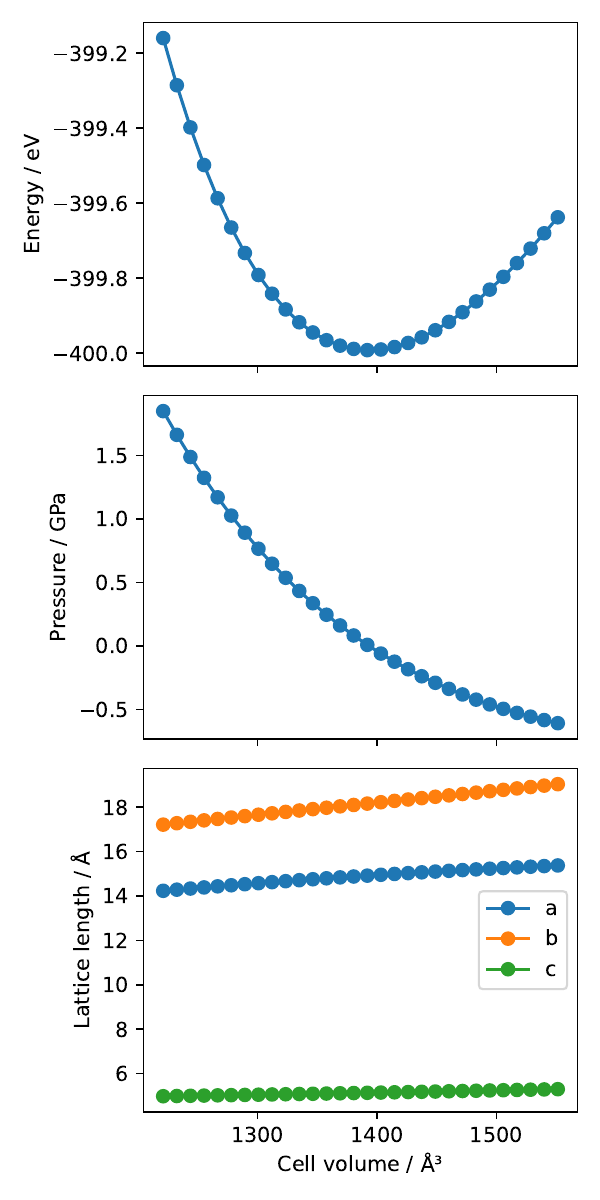}
 \caption{Lattice properties as a function of cell volume. Top: lattice energy, middle:pressure, bottom: individual lattice vector length.}
 \label{eos}
\end{figure}

Additional calculations were performed on the generated MLIP to assess how lattice parameters and lattice energy evolves as a function of pressure.
Energy--volume curves were produced using \codename{janus-core}: at each point the experimental structure is scaled then the lattice parameters and geometry optimised with the MLIP under a fixed-volume constraint.
The Birch--Murnaghan equation of state was fitted to the resulting data and used to obtain a hydrostatic pressure estimate at each point.
The results are contained in Figure~\ref{eos}.

\begin{figure}[htbp]
 \centering
 \includegraphics[width=0.6\linewidth]{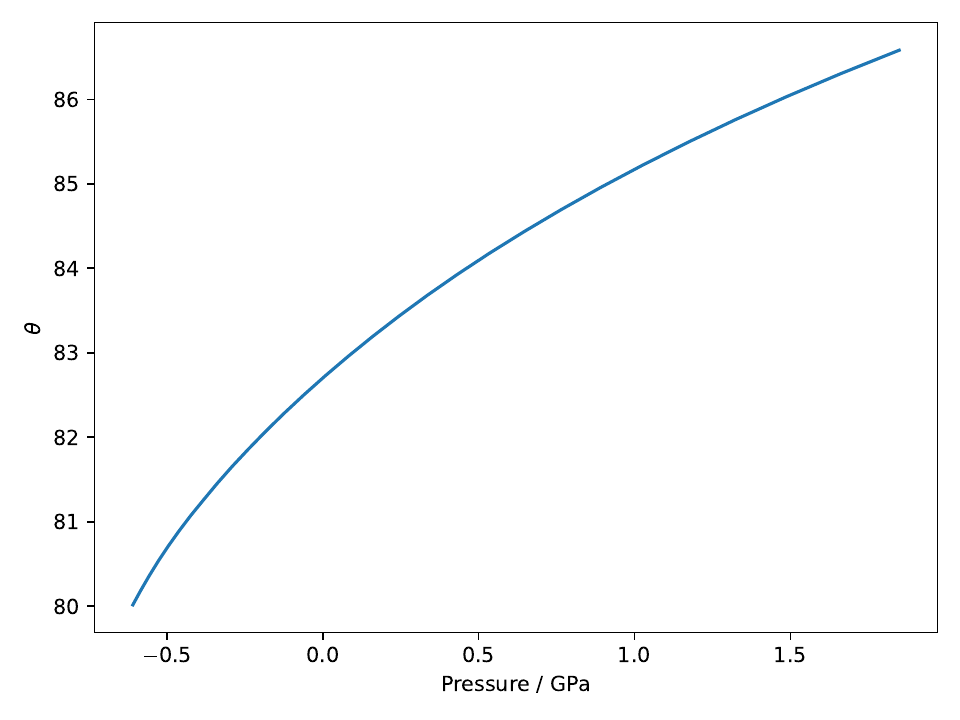}
 \caption{Angle between the ring normals on each side of the ``herringbone'' arrangement of molecules in 2,5-diidothiophene as a function of pressure.
 This was calculated using structures and Birch--Murnaghan equation of state from MLIP energy-volume curve (Figure~\ref{eos}).}
 \label{angles}
\end{figure}

At each volume the rings were identified and isolated using neighbour-lists in ASE, and their out-of-plane (normal) vectors identified by principle-component analysis in scikit-learn.\cite{hjorthlarsenAtomicSimulationEnvironment2017,scikit-learn}
The vectors were split into two groups (corresponding to each side of a "herringbone" row of molecules in 2,5-diidothiophene) and a representative angle calculated between their averages.
The evolution of this angle is plotted against the corresponding EOS pressure in Figure~\ref{angles}.

% ============================================================
\FloatBarrier
\printbibliography